\documentclass[a4paper,8pt]{article}
\usepackage[latin1]{inputenc}

\newtheorem{thm}{Théorème}[section]
\newtheorem{prop}[thm]{Proposition}

\newtheorem{resul}[thm]{Résultat}

\newcommand{\calA}{{\cal A}}
\newcommand{\ra}{{\rangle}}
\newcommand{\la}{{\langle}}
\def\dem {\noindent {\bf Proof : }}

\def\sqw{\hbox{\rlap{\leavevmode\raise.3ex\hbox{$\sqcap$}}$\sqcup$}}
\def\findem{\ifmmode\sqw\else{\ifhmode\unskip\fi\nobreak\hfil
\penalty50\hskip1em\null\nobreak\hfil\sqw
\parfillskip=0pt\finalhyphendemerits=0\endgraf}\fi}

\usepackage{latexsym}
\usepackage{amsmath,amssymb,amsfonts,graphicx}

\newcommand{\dsp}{\displaystyle}

\begin{document}

\begin{titlepage}

\begin{flushright}
CPT-P003-2009 \\
Mars 2009
\end{flushright}
\vspace{1.cm}

\begin{center}
{\LARGE\bf Reduction of one-massless-loop with scalar boxes in $n+2$ dimensions.}\\[1cm]%

{C.~Bernicot $^{a}$}\\[1cm]%

{\em $^{a}$ CPT, Luminy, CNRS\\
UMR 6207 \\
Campus de Luminy, Case 907 - 13288 Marseille cedex 9, France.}\\[.5cm]

\end{center}
\normalsize

\vspace{1cm}

\begin{abstract}

All one-massless-loop Feynman diagrams could be written like a
linear combination of scalar boxes, triangles an bubbles in $n$
dimensions plus rational terms. However, the four-point scalar
integrals in $n+2$ dimensions are free of infrared divergences. We
are going to change the dimensions of the scalar boxes $n
\rightarrow n+2$ and the using of this degree of freedom to
simplify the computation of coefficients of the decomposition.

\end{abstract}

\vspace{5cm}

version \today

\end{titlepage}

\section{Introduction}

\hfil

Since many years, one have try to calculate analytically Feynman
Diagrams. The aim is the computation of amplitudes and cross
sections. In fact, at the LHC, we want to discover new physics and
new particles like Higgs by the interaction of two protons. The
background is constituted by many QCD reactions. So the knowlegde
of this pone is necessary, if we want to detect a new particle.
But the amplitudes depend to an unphysical energy, and this
dependance decreases with the order of the development. So to have
a good prediction, we have to calculate each reaction at NLO. At
this time, all $2\rightarrow2$ and $2\rightarrow3$ processes are
known at NLO, but it remains the $2\rightarrow N$, with $ N > 3$.
So we have to calculate one-loop diagrams with many ingoing legs.

\hfil

Forty years ago, Passarino and Veltman gave a first method to
reduce diagrams. This method is not really efficient and doesn't
use the mathematical symmetry of a loop, for example, unitarity.
They showed that we can write a diagram like a linear combination
of scalar integrals. Then since ten years, Bern and al.
\cite{Bern:massiveloop,Dixon:TASI,bern:ee4partons,Bern:rational}
reduce diagrams thanks to unitarity. It simplifies computation. In
2004, Britto and al.
\cite{Britto:2004nc,Britto:2005ha,Britto:2006sj,method7,method7bis}
gave a very efficient method to calculate the coefficients in
front of the four-point scalar integrals. Then, Mastrolia
\cite{Mastrolia:2006ki} found a way to express the coefficients in
front of triangles. Recently Forde \cite{Forde:2007mi} in one hand
and Papadopoulos and al. \cite{Papa:cut} in a other hand gave some
algorithms to obtain directly the coefficients in front of the
scalar massless boxes, triangles and bubbles. Finally, Kilgore
\cite{Kilgore} improves the Forde algorithms to a massive loop.
But in this algorithm, less the scalar integrals have legs, more
the coefficients are difficult to calculate because some free
parameters appear. In fact, as they use unitarity and cuts, we
need four cuts to define a loop momenta. But we have only three
cuts (respectively two cuts) in a triangle (resp. in a bubble). So
some free parameters remains.

\hfil

In this paper, I would like to give a way to eliminate this degree
of freedom in front of the triangles in a massless loop. Just
after given some notations in the section $\ref{notations}$, I
would like to speak about the bases of scalar integrals in section
$\ref{basedecompo}$ and to show that the classical base is not
efficient, therefore I give a better base. Then, in section
$\ref{coefficientf}$, I would like to give a way to find
coefficients in front of triangles in this new bases. Finally I
finish by an exemple in section $\ref{examplee}$: the four-photon
amplitudes. For a massive loop more work are needed.

\hfil

\section{Notations} \label{notations}

\hfil

\begin{figure}[httb!]
\centering
\includegraphics[width=4cm]{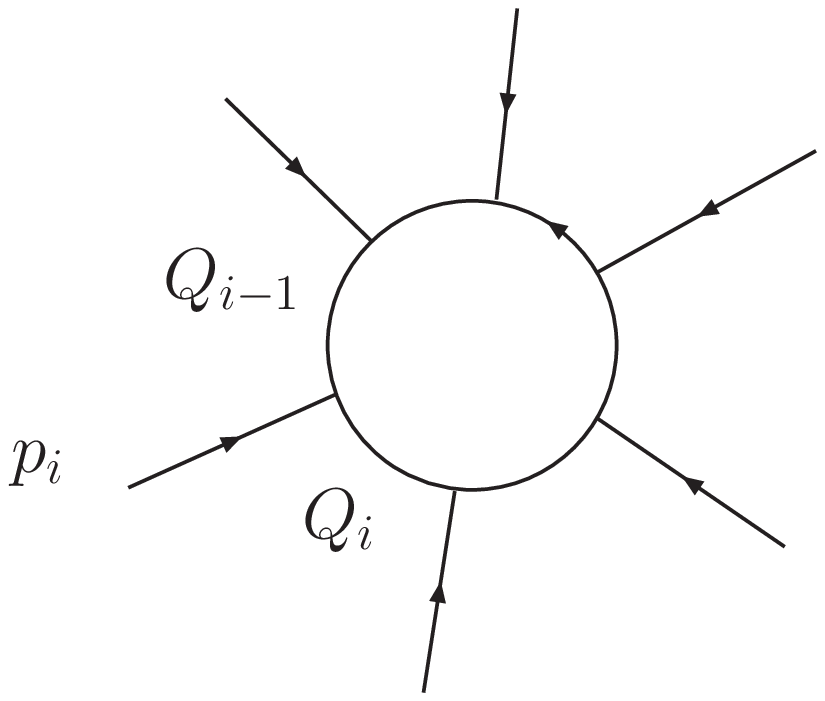}
\caption{\scriptsize \textit{General structure of a
loop.}}\label{structureboucle}
\end{figure}

So in this first section we gave some notations. The purpose of
this article is to study the decomposition of a one-massless loop
Feynman diagram amplitude. So consider a one-loop diagram with $N$
ingoing legs (Fig. $\ref{structureboucle}$), we note it amplitude:
\begin{align}
    {\cal A}_{N} \ = \ \int d^{n} Q
    \frac{\textrm{Num}(Q)}{D_{1}^{2}...D_{N}^{2}},
    \label{ampliini}
\end{align}
\noindent with $Q$ the loop momentum in $n = 4 -2 \epsilon$
dimensions and the denominator $D_{i}^{2} \ = \ Q_{i}^{2} +
i\lambda \ = \ \left( Q + r_{i} \right)^{2} + i \lambda$. We
decide to note the $n$-dimensional vectors in capital letters and
the $4$-dimensional vectors in small letters. The $-2\epsilon$
part of the loop momentum is $\mu$. As the four and the
$-2\epsilon$ parts Minkowski space are orthogonal, therefore the
denominator ``i'' is written:
\begin{equation}
    D_{i}^{2} \ = \ Q_{i}^{2} + i\lambda \ = \ \left( q_{i} + \mu
    \right)^{2} + i \lambda \ = \ q_{i}^{2} + i\lambda - \mu^{2} \ = \
    d_{i}^{2} - \mu^{2}. \label{4DDeno}
\end{equation}
\noindent ``$d_{i}^{2}$'' is the four dimensional part of the
denominator ``i''. The function $\textrm{Num}(Q)$ depends on the
theory with described the loop. In classical gauge theory, this
function is polynomial.

\hfil

In particularly, we note $I_{N}^{n}$ the amplitude of a one-loop
diagram, with $N$ external legs, where all internal propagators
are scalar:
\begin{align}
    I_{N}^{n} \ = \ \int d^{n} Q \frac{1}{D_{1}^{2}...D_{N}^{2}}.
\end{align}
\noindent We often call those amplitudes: scalar integrals. There
are well-known and we can express them with explicit analytic
expressions. We recall some of them in Appendix \ref{scalar_int}.
Moreover, we use oftentimes the kinematical matrix ${\cal S}_{ij}$
and the Gram matrix $G_{ij}$, defined (for a massless loop) by:
\begin{align}
    {\cal S}_{ij} \ = \ \left( q_{i}-q_{j}\right)^{2}  \quad \quad
    G_{ij} \ = \ 2 p_{i}.p_{j}
\end{align}

\hfil

\section{The bases of decomposition} \label{basedecompo}

\hfil

Now we assume a one-massless-loop Feynman diagram, described in a
gauge theory by the amplitude $\left( \ref{ampliini} \right)$. The
numerator of the amplitude is a polynomial function. This last
point out is very importante to decompose the amplitude.

\hfil

\subsection{The classical base}

\hfil

As the loop are described in a standard gauge theory, therefore,
the integrand of the amplitude is a rational function. We can
expand it automatically in partial fractions. Each one gives a
scalar integral. Therefore, we can write straightforward an
amplitude like a linear combination of scalar integrals in $n$
dimensions:
\begin{align}
    {\cal A}_{N} \ = \ \sum_{i=1}^{N} a_{i} I_{i}^{n} + \textrm{Rationnal terms} + O(\epsilon).
\end{align}
\noindent Obviously the decomposition of an $N$ external-leg
one-loop Feynman diagram uses scalar integrals with at most $N$
external legs. However, all those integrals $\left\{ I_{i}^{n}, \
n \in [1.. N] \right\}$ are not free. Indeed, we have the linear
relation \cite{reduction,method1}:
\begin{equation}
        I_{N}^{n}\left( {\cal S} \right) \ = \ \sum_{i,j=1}^{N} {\cal S}_{ij}^{-1} {I_{N-1}^{n}\left( {\cal S}-\{i\} \right)}_{i}
        - (-1)^{N+1} \left( N-n-1 \right) \frac{\det \left(G\right)}{\det \left( {\cal S} \right) }I_{N}^{n+2}\left( {\cal S}
        \right), \label{reductionint}
\end{equation}
\noindent where ${\cal S}-\{i\}$ is the kinematical matrix
obtained by eliminate the column and the line number ``i''.
Moreover, we can show that the Gram determinant for $N> 5$ is
equal to zero. So this last equation $\left( \ref{reductionint}
\right)$ simplify:
\begin{align}
    \left\{ \begin{array}{l}
    \dsp \forall \ N > 5, \ I_{N}^{n}\left( {\cal S} \right) \ = \ \sum_{i,j=1}^{N} {\cal S}_{ij}^{-1} {I_{N-1}^{n}\left( {\cal S}-\{i\}
    \right)}_{i} \\
    \dsp I_{5}^{n}\left( {\cal S} \right) \ = \ \sum_{i,j=1}^{5} {\cal S}_{ij}^{-1} {I_{4}^{n}\left( {\cal S}-\{i\} \right)}_{i}
    - 2 \epsilon \ \frac{\det \left(G\right)}{\det \left( {\cal S} \right) }I_{5}^{n+2}\left( {\cal S}
    \right).
    \end{array} \right.
\end{align}
\noindent So we see easily that in the leading order in
$\epsilon$, the reduction $\left( \ref{reductionint} \right)$
simplify (as the loop is massless, we don't need tadpoles to
decompose the loop):
\begin{align}
    {\cal A}_{N} \ = \ \sum_{i=2}^{4} a_{i} I_{i}^{n} +
    O(\epsilon). \label{reduction}
\end{align}
\noindent We need only scalar boxes, triangles, bubbles and
rational terms to generate all the one-massless-loop Feynman
diagrams. Generally we call the scalar function part: the analytic
part and the rational terms the rational part. The call ``analytic
part'' comes from fact that they contains polylogarithm functions
of Mandelstam variables. The rational terms are studied in
\cite{Papadopoulos:rational,binoth2}. But we are going to see,
that this base ${\cal B} \ = \ \{ I_{2}^{n},...I_{4}^{n} \}$ is
not efficient.
\begin{align*}
    {\cal A}_{N} \ = \ \sum_{i} \ &
    a_{i} \parbox{2cm}{\includegraphics[width=2cm]{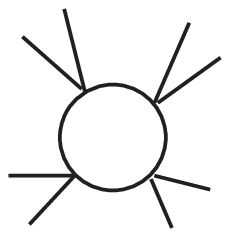}}^{n}+
    b_{i} \parbox{2cm}{\includegraphics[width=2cm]{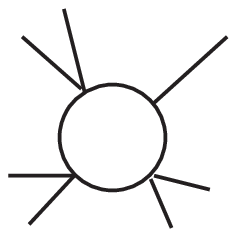}}^{n}+
    c_{i} \parbox{2cm}{\includegraphics[width=2cm]{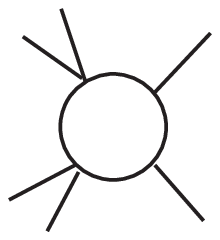}}^{n}+
    & \\
    + & d_{i}\parbox{2cm}{\includegraphics[width=2cm]{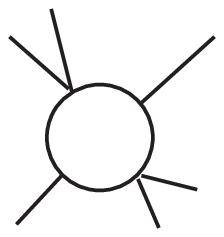}}^{n}+
    e_{i}\parbox{2cm}{\includegraphics[width=2cm]{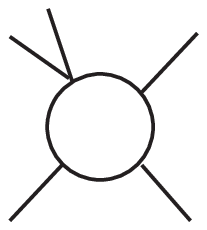}}^{n}+
    f_{i}\parbox{2cm}{\includegraphics[width=2cm]{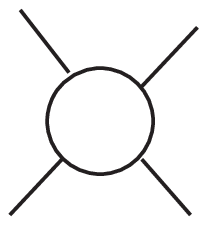}}^{n}+ & \\
    + & g_{i}\parbox{2cm}{\includegraphics[width=2cm]{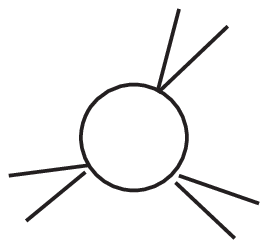}}^{n}+
    h_{i}\parbox{2cm}{\includegraphics[width=2cm]{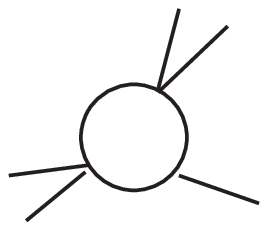}}^{n}+
    i_{i}\parbox{2cm}{\includegraphics[width=2cm]{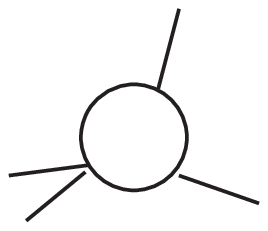}}^{n}+ & \\
    + & j_{i}\parbox{2cm}{\includegraphics[width=2cm]{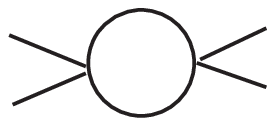}}^{n}+ \textrm{Rational Terms}
\end{align*}
\noindent The acknowledge of the coefficients $\{ a_{i}, ... j_{i}
\}$ and the rational terms is enough to reconstruction all the
amplitude. This base ${\cal B} \ = \ \{ I_{2}^{n},...I_{4}^{n} \}$
is the most obvious, the canonical base, but not the more
efficient. It is the subject of the next subsection.

\hfil

\subsection{The problems of this base $ {\cal B}$}

\hfil

The base of decomposition ${\cal B} \ = \ \{
I_{2}^{n},...I_{4}^{n} \}$ has at least two problems.

\hfil

The first comes from divergences. Indeed, the four-point functions
are constituted by finite terms and infrared divergent terms, the
three-point functions are infrared divergent and the bubbles: UV
divergent. So three and four-point functions could have infrared
divergences. Therefore, with the base ${\cal B}$, we don't have
separate explicitly the infrared, ultraviolet and finite
structure. It could forget some compensations and it is not easy
to found them. For instance, consider a loop without infrared
divergences, therefore we can have three and four-point functions
and not explicite compensations, this implies some numerical
instabilities.

\hfil

The second problem concerns the famous Gram determinant. The
reduction gives some negative powers of Gram determinants in front
of the four-point functions. This determinant is spurious, because
it could be equal to zero but this divergence is not physic, it
has no physical explanation. Indeed, there is some unexplicit
compensations to eliminate those determinants.

\hfil

Those two problems give the base ${\cal B}$ not efficient. So we
have to find an other base.

\hfil

\subsection{A better base}

\hfil

The main problem is the blend of infrared, ultraviolet and finite
parts and in particularly, it should be genius if we could
separate the infrared and the finite part of the four-point scalar
functions. We could do it just by the application of the formula
$\left( \ref{reductionint} \right)$ to the four-point scalar
integrals:
\begin{equation}
        I_{4}^{n}\left( {\cal S} \right) \ = \ \sum_{i,j=1}^{4} {\cal S}_{ij}^{-1} {I_{3}^{n}\left( {\cal S}-\{i\} \right)}_{i}
        -  \left( 1+2\epsilon \right) \frac{\det \left(G\right)}{\det \left( {\cal S} \right) }I_{4}^{n+2}\left( {\cal
        S} \right) \label{equredcution}.
\end{equation}
\noindent The four-point scalar functions in $n+2$ dimensions are
totally finite and well-known, we recall them in Appendix
\ref{scalar_int}. Moreover $\left( \ref{equredcution} \right)$ see
us that it appears one Gram determinant in the numerator of the
coefficients in front of the four-point functions. So with $\left(
\ref{equredcution} \right)$, we reduce one time the problem of the
Gram determinant, but we don't eliminate all the problem. So we
conclude that the base ${\cal B}^{'} \ = \ \left\{
I_{2}^{n},I_{3}^{n},I_{4}^{n+2} \right\}$ is better than ${\cal
B}$. It remains to calculate the coefficients in front of each
scalar integral.
\begin{align*}
    {\cal A}_{N} \ = \ \sum_{i} \ &
    a_{i}^{'} \parbox{2cm}{\includegraphics[width=2cm]{I4m4.eps}}^{n+2}+
    b_{i}^{'} \parbox{2cm}{\includegraphics[width=2cm]{I4m3.eps}}^{n+2}+
    c_{i}^{'} \parbox{2cm}{\includegraphics[width=2cm]{I4m2A.eps}}^{n+2}+
    & \\
    + & d_{i}^{'}\parbox{2cm}{\includegraphics[width=2cm]{I4m2B.eps}}^{n+2}+
    e_{i}^{'}\parbox{2cm}{\includegraphics[width=2cm]{I4m1.eps}}^{n+2}+
    f_{i}^{'}\parbox{2cm}{\includegraphics[width=2cm]{I4m0.eps}}^{n+2}+ & \\
    + & g_{i}^{'}\parbox{2cm}{\includegraphics[width=2cm]{I3m3.eps}}^{n}+
    h_{i}^{'}\parbox{2cm}{\includegraphics[width=2cm]{I3m2.eps}}^{n}+
    i_{i}^{'}\parbox{2cm}{\includegraphics[width=2cm]{I3m1.eps}}^{n}+ & \\
    + & j_{i}\parbox{2cm}{\includegraphics[width=2cm]{I2m1.eps}}^{n}+ \textrm{Rationnal
    Terms}.
\end{align*}
\noindent The ultraviolet divergences are carried out by the
scalar bubbles, whereas the infrared structures are contained by
the one and two-external-mass scalar triangles, but the finite
structure are the four-point functions and the three-external-mass
scalar triangle. So this decomposition partitions the amplitudes
over the three kind of analytic structure without covering.

\hfil

We will see that this partition is very interesting. It eliminates
many subtil compensations. And the result that we are going to
show is that if the loop has no infrared divergences, so the
decomposition has no infrared structure. The extension of this
result to the ultraviolet one is not so easy.

\hfil

\section{Coefficients in front of boxes and triangles in the base ${\cal B}^{'} = \left\{
I_{2}^{n},I_{3}^{n},I_{4}^{n+2} \right\}$} \label{coefficientf}

\hfil

Now we are going to calculate the coefficients $\{
a_{i}^{'},...,j_{i} \}$ in this new base ${\cal B}^{'} = \left\{
I_{2}^{n},I_{3}^{n},I_{4}^{n+2} \right\}$. In this section, we
give the method to calculate the coefficients of the four and
three-point functions. The ones in front of bubbles will be given
in an other paper.

\hfil

\subsection{Coefficients of four-point functions.}

\hfil

To transform the base ${\cal B} \ = \ \left\{
I_{2}^{n},I_{3}^{n},I_{4}^{n} \right\}$ to the base ${\cal B}^{'}\
= \ \left\{ I_{2}^{n},I_{3}^{n},I_{4}^{n+2} \right\}$, we have
just to apply the formula $\left( \ref{equredcution} \right)$ to
the first base. So keeping only the four-point scalar function
part, the transformation gives us:
\begin{equation}
    a_{i} I_{4}^{n} \ \rightarrow \ -  \left( 4-n-1 \right) \frac{\det \left(G\right)}{\det \left( {\cal S} \right) }
    a_{i}I_{4}^{n+2}.
\end{equation}
\noindent So, we obtain the bounding equation:
\begin{equation}
    a_{i}^{'} \ = \ -  \left( 4-n-1 \right) \frac{\det \left(G\right)}{\det \left( {\cal S} \right) }
    a_{i}, \label{coef4}
\end{equation}
\noindent where $G$ (respectively $\cal S$) are the Gram matrix
(resp. kinematical matrix) of the four-point scalar function. The
coefficient $a_{i}$ is given directly by the unitarity-cut method
\cite{Britto:2006sj,Cutkosky}. We recall it, for example, consider
a loop with $N$ ingoing legs described by the amplitude $\left(
\ref{ampliini} \right)$ and, we want the coefficient in front of
the box obtained by pinching the denominators of the initial loop
except the four ones called ``$a,b,c$'' and ``$d$''. Therefore,
the unitary cuts give us directly the coefficient by the limit:
\begin{equation}
    a_{i} \ = \ \lim_{d_{a}^{2},d_{b}^{2},d_{c}^{2},d_{d}^{2} \rightarrow 0} \frac{\textrm{Num}\left(q \right)
    d_{a}^{2}d_{b}^{2}d_{c}^{2}d_{d}^{2}}{d_{1}^{2}...d_{N}^{2}},
\end{equation}
\noindent where $d_{i}^{2}$ represents the four-dimensional part
of the denominator $D_{i}^{2}$: $d_{i}^{2} = q_{i}^{2} + i
\lambda$ (eq. $\ref{4DDeno}$). To obtain this limit we have just
to solve the system $\{ i \in [a..d], d_{i}^{2} =0 \}$. The loop
momentum is a four-dimensional vector, so we write it like a
linear combination of four four-dimensional vectors (for example
external ingoing legs). The system is linear, with four equations
and four variables (four parameters of the linear combination of
the loop momentum), so we can solve it exactly, and we explicite
the momentum of the loop. We note $q_{0}$ this momentum. This
process is explained in
\cite{Britto:2004nc,Britto:2006sj,Papa:cut,method8,Bernicot4g}.

\hfil

\subsection{Coefficients of three-point functions.}

\hfil

In this subsection we use the Forde results and the Forde
formalism given in \cite{Forde:2007mi}. Here we give the method to
find the coefficient in front the three-point functions, assuming
that the four-point functions are in $n+2$ dimensions. To simplify
the proof, we use an amplitude with four external legs:
\begin{equation}
    \calA_{4} \ = \ \int d^{n} Q \frac{\textrm{Num}\left( Q \right)}{D_{1}^{2}D_{2}^{2}D_{3}^{2}D_{4}^{2}},
\end{equation}
\noindent which we decompose on the base ${\cal B}^{'}= \left\{
I_{2}^{n},I_{3}^{n},I_{4}^{n+2} \right\}$ and rational terms:
\begin{equation}
    {\cal A}_{4} \ = \  - a \frac{\det \left( G \right)}{\det \left( {\cal S} \right)} I_{4}^{n+2} + \sum_{i=1}^{4} T_{i} I_{3,i}^{n}
    + \sum_{i=1}^{2} \gamma_{i} I_{2}^{n} + \textrm{rationnal terms}.
\end{equation}
\noindent With the last subsection, we can compute, very easily
the coefficient in front of the four-point function, we know $a =
\textrm{Num}(q_{0})$, where $q_{0}$ is the solve of the linear
system $\left\{ \forall i \in[1..4], \ d_{i}^{2} =0 \right.$ given
by the four cuts. Now we want to calculate the coefficients in
front triangles. We are going to use the fact that
\cite{reduction,method1}:
\begin{equation}
    - \frac{ \det \left( G \right)}{\det \left( {\cal S} \right)}
    I_{4}^{n+2} \ = \ \int d^{n} Q \frac{1 -\sum_{i=1}^{4} b_{i}D_{i}^{2}}{D_{1}^{2}D_{2}^{2}D_{3}^{2}D_{4}^{2}},
\end{equation}
\noindent where $b_{i} = \sum_{j=1}^{4} {\cal S}_{ij}^{-1}$. The
amplitude becomes:
\begin{align}
    {\cal A}_{4} \ = \ \int d^{n} Q \frac{\textrm{Num}\left( Q \right)}{D_{1}^{2}D_{2}^{2}D_{3}^{2}D_{4}^{2}} \ = & \ \textrm{Num}
    \left( q_{0} \right) \int d^{n} Q \frac{1 -\sum_{i=1}^{4}
    b_{i}D_{i}^{2}}{D_{1}^{2}D_{2}^{2}D_{3}^{2}D_{4}^{2}} & \nonumber \\
    & + \sum_{i=1}^{4} T_{i} \int d^{n}Q
    \frac{D_{i}^{2}}{D_{1}^{2}D_{2}^{2}D_{3}^{2}D_{4}^{2}}
    + \sum_{i=1}^{2} \gamma_{i} I_{2}^{n} + \textrm{Rationnal Terms}. \label{III} &
\end{align}
\noindent For example, we assume that we want to calculate the
coefficient in front of the three-point scalar function obtained
by pinching the propagator number 1. So we apply, in the last
equation $\left( \ref{III} \right)$, the linear application
``$\textrm{Disc}_{2,3,4}$'', which cuts the three propagators 2,3
and 4 in four dimensions:
\begin{align}
    \textrm{Disc}_{2,3,4} \calA _{4} \ = \ \textrm{Disc}_{2,3,4} \int d^{n} Q \frac{\textrm{Num}\left( Q \right)}{D_{1}^{2}D_{2}^{2}D_{3}^{2}D_{4}^{2}} \ = & \ \textrm{Num}
    \left( q_{0} \right) \textrm{Disc}_{2,3,4} \int d^{n} Q \frac{1 -\sum_{i=1}^{4}
    b_{i}D_{i}^{2}}{D_{1}^{2}D_{2}^{2}D_{3}^{2}D_{4}^{2}} & \nonumber \\
    & + \sum_{i=1}^{4} T_{i} \ \textrm{Disc}_{2,3,4} \int d^{n}Q
    \frac{D_{i}^{2}}{D_{1}^{2}D_{2}^{2}D_{3}^{2}D_{4}^{2}} \label{IIII}.
\end{align}
\noindent We keep on only the three and four-point functions. The
application ``$\textrm{Disc}_{2,3,4}$'' inputs the three
propagators $d_{2}^{2},d_{3}^{2}$ and $d_{4}^{2}$ on-shell. As we
keep only the four-dimensional part in the numerator, because we
want the coefficient in front of the scalar integrals. Indeed, the
$-2\epsilon$ dimensional part of the loop momentum in the
numerator give the rational terms \cite{Papadopoulos:rational}.
Therefore $\left( \ref{IIII} \right)$ becomes:
\begin{align}
   \int d^{n} Q \ \frac{\textrm{Num}\left( Q \right)}{D_{1}^{2}} \ \delta\left(2,3,4 \right) \ = & \ \textrm{Num}
    \left( q_{0} \right) \int d^{n} Q \frac{1 -b_{1}D_{1}^{2}}{D_{1}^{2}} \delta\left(2,3,4 \right) + T_{1} \int d^{n}Q \ \delta\left(2,3,4 \right),
\end{align}
\noindent where $\delta\left(2,3,4 \right) = \delta\left(
d_{2}^{2}\right)\delta\left( d_{3}^{2}\right)\delta\left(
d_{4}^{2}\right)$. In this step we use the Forde Formalism
\cite{Forde:2007mi}. As we have only three cuts, we have only
three equations. We write again the loop momentum like a linear
combination of four four-dimensional vectors $\left\{
{K_{3}^{b}}^{\mu},{K_{4}^{b}}^{\mu}, \la K_{3}^{b} \gamma^{\mu}
K_{4}^{b} \ra, \la K_{4}^{b} \gamma^{\mu} K_{3}^{b} \ra \right\}$,
where ${K_{3}^{b}}^{\mu}$ and ${K_{4}^{b}}^{\mu}$ are light-like
vectors defined in the Appendix $\ref{Forderecall}$. But as we
have only three equations, therefore, we can explicit only three
parameters over four of the linear combination of the loop
momentum. The one, which it remains, is noted ``c''. According to
\cite{Forde:2007mi}, $q_{3}$ becomes:
\begin{equation}
    q_{3}(c) \ = \ \alpha_{04} {K_{3}^{b}}^{\mu} + \alpha_{03} {K_{4}^{b}}^{\mu} + \frac{c}{2} \la K_{3}^{b} \gamma^{\mu}
    K_{4}^{b} \ra + \frac{\alpha_{03}\alpha_{04}}{2c} \la K_{4}^{b} \gamma^{\mu} K_{3}^{b}
    \ra .
\end{equation}
\noindent where all $\alpha_{ij}$ and vectors $K_{i}$ are
explained in Forde paper \cite{Forde:2007mi} and recalled in the
Appendix $\ref{Forderecall}$. Moreover, we use the spinor
notations introduced in \cite{spinor:chinois}. Now we change the
variables of integration and do the integrations over the three
delta functions. We obtain:
\begin{equation}
    \ \int dc \ J_{c} \ \frac{\textrm{Num}\left( c \right)}{D_{1}(c)^{2}} \ = \ \textrm{Num}
    \left( q_{0} \right) \int dc \ J_{c} \frac{1 - b_{1}D_{1}^{2}(c)}{D_{1}^{2}(c)} + T_{1} \int dc \
    J_{c}.\label{OPQ}
\end{equation}
\noindent If we want to calculate the coefficient of the three
point function, we have just to solve the equation:
\begin{equation}
    1 - b_{1}d_{1}^{2} \left( c_{0} \right)\ = \ 0, \label{equationXXX}
\end{equation}
\noindent where $D_{1}^{2} = d_{1}^{2} - \mu^{2}$ (eq.
$\ref{4DDeno}$). We put those $I$ solves $c_{0}^{(i)}$ in the
equation $\left( \ref{OPQ} \right)$ and we obtain the result.

\begin{resul}
     In the base ${\cal B}^{'}$, the coefficient in front of the scalar integrals $I_{3,1}^{n}$ is:
     \begin{equation}
        T_{1} \ = \ b_{1} \sum_{i=1}^{I} \textrm{Num}\left(
        c_{0}^{(i)} \right),
     \end{equation}
     \noindent where $c_{0}$ is the solve of the equation $\left( \ref{equationXXX}
     \right)$. For the other triangles, we have just to permute the cut propagators and the parameter $b_{i}$.
\end{resul}

\hfil

Here we see one of the interest of the base ${\cal B}^{'}$ rather
than the base ${\cal B}$. Indeed, not only the base ${\cal B}^{'}$
separate the infrared, ultraviolet and the finite parts but also
the coefficients in front of scalar triangles are very simple to
computed. There are no longer problems to obtain the free
parameters, it is given by the equation $\left( \ref{equationXXX}
\right)$. The solve of this equation is often obvious. In the last
example, if one of the legs $p_{3}$ or $p_{4}$ is massless
therefore, the solve is zero: $c_{0} =0$.

\hfil

To improve the computation in this base ${\cal B}^{'}$, we are
going to give some rules. With them, we are going to know
directly, without computation, the null coefficients in front of
triangles.

\hfil

\subsection{Infrared Divergences}

\hfil

\begin{prop} \label{existenceintegralinfrarouge}
    Consider the decomposition of an one-massless-loop Feynman diagram on the base
    ${\cal B}^{'} \ = \ \left\{ I_{2}^{n},I_{3}^{n},I_{4}^{n+2} \right\}$.
    We assume that this diagram has no infrared (soft or collinear) divergence. Therefore the coefficients in front of the
    scalar one or two-external-massive-leg triangles are zero.
\end{prop}

\dem We can give an analytical proof, but with some arguments of
reduction, the proof is obvious.\\
Consider a one-loop-massless diagram free of IR divergences. The
reduction by standard methods gives some sub-diagrams by pinching
propagators. But those reduction cannot create infrared
divergences. Indeed the pinched propagator plus the two-adjacent
legs give a massive-external leg, which eliminates all infrared
divergences around it. After the reduction, we obtain three-point
sub-diagrams and the ``finite'' part of four-point scalar
integrals. Since the three-point sub-diagrams are free of IR
divergences, they cannot be expressed in term of one mass/two mass
three-point scalar integrals and so the coefficients in front of
them are zero. \findem

\hfil

We can improve this result and consider a diagram with infrared
divergences.

\begin{prop} \label{existenceintegralinfrarouge2}
    Consider the decomposition of an one-massless-loop Feynman diagram, with a soft divergence on the propagator number ``i'',
    on the base ${\cal B}^{'} \ = \ \left\{ I_{2}^{n},I_{3}^{n},I_{4}^{n+2} \right\}$. The coefficients in
    front of the one-external and two-external-massive-leg triangles are
    zero except the one which corresponds to the one-external-massive-leg triangle whom
    the external-massive-leg is opposite to the propagator number ``i'':
    \begin{equation*}
        \parbox{2cm}{\includegraphics[width=2cm]{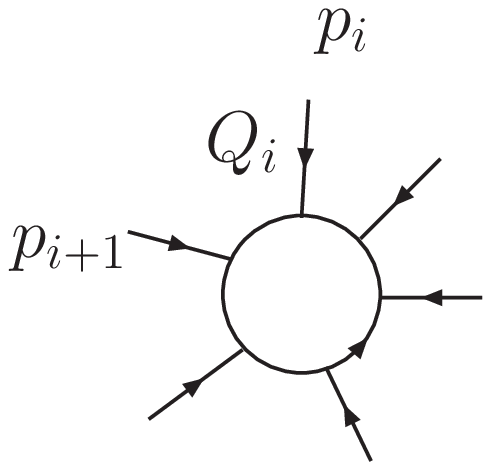}}
        =  \sum_{i} a_{i} {I_{4}^{n+2}}_{i} + b \
        \left(\parbox{2cm}{\includegraphics[width=2cm]{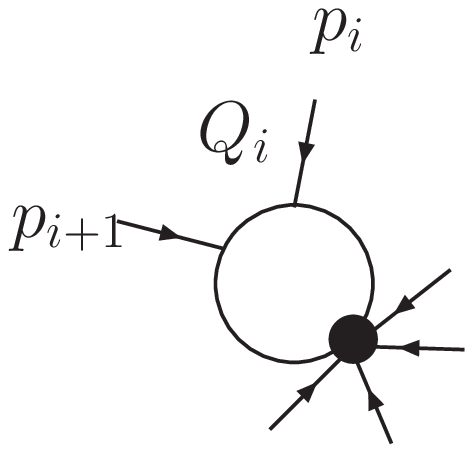}}\right)+
        \sum_{i} c_{i} {I_{2}^{n}}_{i}
    \end{equation*}
\end{prop}

\begin{prop} \label{existenceintegralinfrarouge3}
    Consider the decomposition of a one-massless-loop Feynman diagram, with a collinear divergence on the external leg ``i'',
    on the base $\left\{ I_{2}^{n},I_{3}^{n},I_{4}^{n+2} \right\}$. The coefficients in front the one-external and two-external-massive-leg triangles are
    zero except the ones which correspond to the triangles, whom the leg
    number ``i'' doesn't belong to an external mass:
    \begin{equation*}
        \parbox{2cm}{\includegraphics[width=2cm]{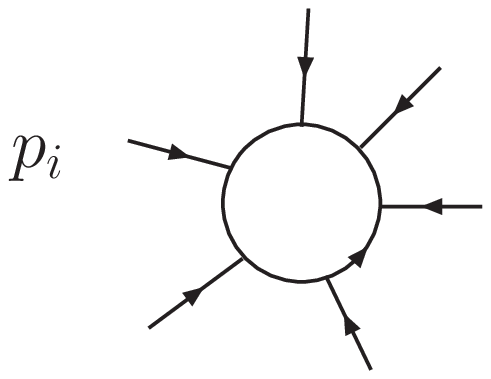}}
        =  \sum_{i} a_{i} \ {I_{4}^{n+2}}_{i} + \sum_{j} b_{j} \
        \left(\parbox{2cm}{\includegraphics[width=2cm]{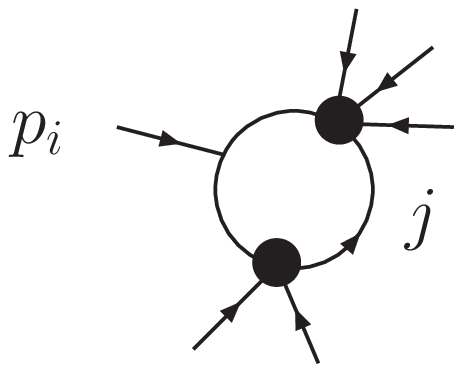}}\right)_{j}+
        \sum_{i} c_{i} \ {I_{2}^{n}}_{i}
    \end{equation*}
\end{prop}

\dem The proofs are the same like the first proposition
$\ref{existenceintegralinfrarouge}$. The external mass regular the
infrared divergences. Therefore all three-point sub-diagrams have
a null coefficient except which one preserves the divergence.
\findem

\hfil

{\bf \noindent Remark:} In a loop, only photons, gluons or scalars
create some divergent propagators. Indeed a fermion propagator has
a numerator and this numerator compensate all infrared
divergences.

\hfil

If we decide to decompose a one-massless-loop Feynman diagram on
the base ${\cal B}^{'} \ = \ \left\{
I_{2}^{n},I_{3}^{n},I_{4}^{n+2} \right\}$, using this last remark
we deduce straightforward the three-point scalar integrals which
have a null or non-null coefficients.

\hfil

\section{An Exemple: the four-photon amplitudes} \label{examplee}

\hfil

We are going to calculate rapidly the helicity amplitudes of the
four-photon amplitudes in QED for example, and in QED, scalar QED
and supersymmetric QED $^{\cal N}=1$. All the result are given in
\cite{karplus,tollis,gamtwo,gamsusy,mahlon2,nagy,papadopoulos:6photon,binoth:2007,bernicot,bernicot1},
but we are going to found the results.

\hfil

The four-photon amplitudes $\gamma_{1}+ \gamma_{2} + \gamma_{3} +
\gamma_{4} \rightarrow 0$ in QED are null at tree order. The first
non null order is the one-loop order. Therefore, the amplitude is
six one-massless-loop diagrams with four photons ingoing in a loop
of fermions. So the decomposition of the amplitude on the base
${\cal B}^{'}$ is:

\begin{equation}
    {\cal A}_{4} \ = \ \sum_{\sigma(1,2,3,4)} \  \left( a^{'} I_{4}^{n+2}
    \left( \sigma \right) + b^{'} I_{3}^{n}\left( \sigma \right) + c I_{2}^{n} \left( \sigma \right) \right) +
    \textrm{rationnal terms}.
\end{equation}
\noindent However with the last subsection, as the loop is a
fermion loop, therefore, there are no infrared divergence so
$b=0$. The decomposition becomes:
\begin{equation}
    {\cal A}_{4} \ = \ \sum_{\sigma(1,2,3,4)} \  \left( a^{'} I_{4}^{n+2}
    \left( \sigma \right) + c I_{2}^{n} \left( \sigma \right) \right) +
    \textrm{rationnal terms}.
\end{equation}
\noindent $a^{'}$ is obtained by the equation $\left( \ref{coef4}
\right)$:
\begin{equation}
    a^{'} \ = \ (1-2\epsilon) \frac{\det(G)}{\det({\cal S})} \ a,
\end{equation}
\noindent where ``$a$'' is given by unitary-cut. If all photons
have a positive or a negative helicity therefore, at least two
adjacent photons have the same helicity, and as the loop is
massless therefore the coefficient is null. We explain it in
\cite{Bernicot4g}. We have the same argument if three photons have
the same helicity:
\begin{equation}
    a(++++) \ = \ a(-+++) \ = \ 0.
\end{equation}.

If we have two positive-helicity photons and two negative-helicity
photons therefore the coefficient is non null only if the
helicities ingoing alternately in the loop. The non null
coefficient corresponds to the scalar integrals $I_{4}^{n+2}
\left( 1^{+},2^{-},3^{+},4^{-} \right)$. The argument in this
integrals gives the order in which the photon ingoing to the loop.
We alternate the negative and the positive helicity photons. With
some calculation with find directly:
\begin{equation}
    a(-+-+) \ = \ - e^{4} \frac{\langle 12 \rangle}{[12]}\frac{[34]}{\langle 34 \rangle} \frac{t^{2}+u^{2}}{s}
\end{equation}

\noindent To compute the rational terms and the coefficients of
bubbles we need other methods. In
\cite{gamtwo,gamsusy,mahlon2,nagy,papadopoulos:6photon,binoth:2007,bernicot,bernicot1},
we find:
\begin{align}
    \dsp {\cal A}_{4}(++++) \ & = \ \dsp 8 i \alpha^{2} \frac{[12][34]}{\langle 12 \rangle \langle 34 \rangle} + O(\epsilon), &\\
    \dsp {\cal A}_{4}(-+++) \ & = \ \dsp 8 i \alpha^{2} \frac{[34] [231]}{ \langle 34 \rangle \langle 231 \rangle} + O(\epsilon), & \\
    \dsp {\cal A}_{4}(--++) \ & = \ \dsp -8 i \alpha^{2} \frac{\langle 12 \rangle}{[12]}\frac{[34]}{\langle 34 \rangle}
    \left\{ 1 + \frac{t^{2}+u^{2}}{s} I_{4}^{n+2}(1324) +  \frac{t-u}{s}\left( I_{2}^{n}(u)-I_{2}^{n}(t) \right) \right\} + O\left( \epsilon \right).
\end{align}
\noindent The definition of each scalar integrals are given in
Appendix $\ref{scalar_int}$. In the last helicity, the
coefficients in front of bubbles are non null. They carried out
ultraviolet divergences, but the amplitude is free of those
divergence. But the soustraction of two bubbles compensate and
those divergence disappear. We can extend those results to the
six-photon amplitudes. All the results are given in
\cite{bernicot,bernicot1}.

\hfil

\section{Conclusion}

\hfil

In this paper, we suggest a way to decompose a one-massless-loop
Feynman diagram.

\hfil

As the other methods of reduction, we decompose an amplitude like
a linear combination of scalar integrals. But contrary to the
methods already existing, here we use the scalar boxes in $n+2$
dimensions. Thanks to this transformation, we win a degree of
freedom which simplify the computation of the coefficients in
front of the scalar triangles.

\hfil

But it remains works to finish the method. Indeed, we have to
incorpore the computation of the coefficients in front of bubbles
and rational terms. But as the example of the four-photon
amplitudes shows, the bound between the amplitude and the
ultraviolet divergences is not so simple as the one with the
infrared divergences.

\hfil

\section*{Acknowledgements}

\hfil

I want to thank T. Binoth, G. Heinrich and P. Mastrolia for
collaboration during the early stages of this work and for useful
discussions. I would like to thank very much the CPT (Centre de
Physique Théorique, Luminy, France) to welcome me during this
year.

\hfil

\begin{appendix}
\renewcommand{\theequation}{\Alph{section}.\arabic{equation}}
\setcounter{equation}{0}

\section{Scalar integrals } \label{scalar_int}

In this appendix, for sake of completeness, the definition of
master massless integrals used in this paper is recalled, more
details can be found in \cite{Binoth:2001vm}. We also give
$\det(G)$ the determinant of the Gram matrix $G_{ij} = 2
p_{i}.p_{j}$ built with the external four momentum and $\det(S)$
the determinant of the kinematical S-matrix defined by $S_{ij} =
\left( q_{j} - q_{i} \right)^{2}$ where the $q_i$ are the four
dimensional momentum flowing in the propagators. All was explain
in section $\ref{notations}$, but we remember the draw.
\begin{figure}[httb!]
\centering
\includegraphics[width=4cm]{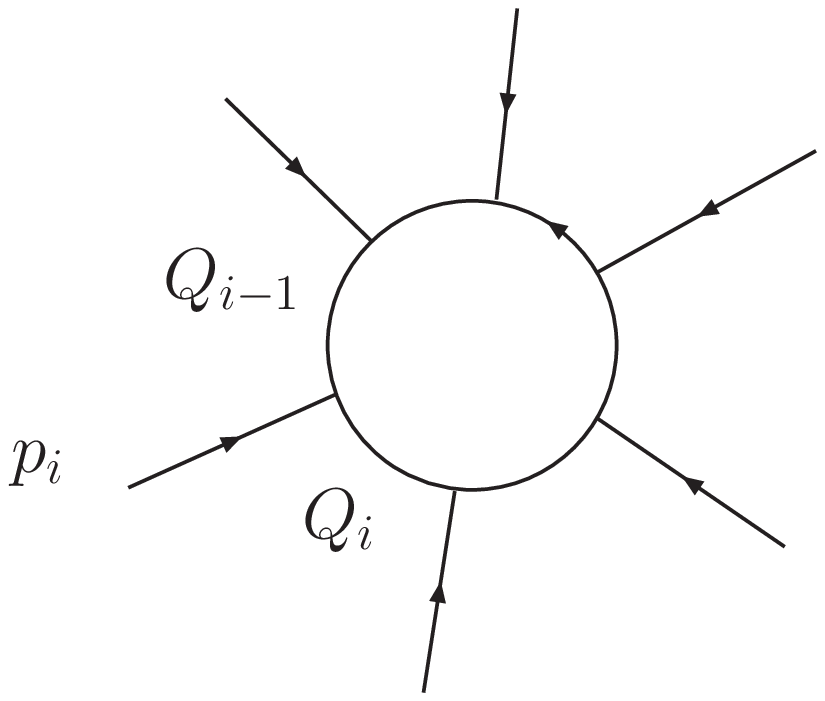}
\caption{\scriptsize \textit{General structure of a loop.}}
\end{figure}
\noindent Moreover, we are going to use the dilogarithm function
defined by \cite{Maximon}:
\begin{equation}
    \textrm{Li}_{2}(x) \ = \ - \int_{0}^{1}dt \  \frac{\ln(1-xt)}{t}.
\end{equation}
\noindent In \cite{Maximon}, we find many formulae using those
dilogaritms. We have to be careful, the Mandelstam variables could
be negative. They are in the arguments of the dilogarithm, so to
solve the problem of the Riemann sheet of the logarithm we use the
analytic continuation by adding a small imaginary part, the
prescription is:
\begin{equation}
    s \rightarrow s + i\lambda, \textrm{with} \ \lambda > 0.
\end{equation}
\noindent And, all the scalar integrals must be multiplied by the
angular integral:
\begin{equation}
    r_{\Gamma} \ = \frac{\Gamma (1 + \epsilon ) \Gamma (1 - \epsilon )^{2}}{\Gamma (1 -2\epsilon
    )}.
\end{equation}
\hfil

\subsection{Two-point Function.}

\begin{align*}
    \parbox{2.5cm}{\includegraphics[width=2.5cm]{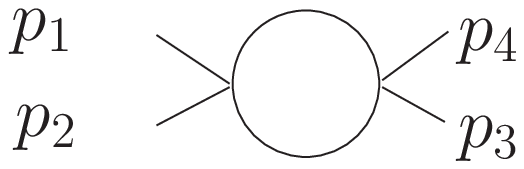}}
    \quad \quad \quad \quad \quad \quad \quad \quad
    s = s_{12}= s_{34}
\end{align*}

\begin{align}
{{\cal S}_{2}} =
\begin{pmatrix}
    0 & s  \\
    s & 0
\end{pmatrix}.
\end{align}

\noindent The determinants are given by:
\begin{align}
    & \det \left( S_{2} \right) \ = \ -s^{2} &\\
    & \det \left( G_{2} \right) \ = \ s.
\end{align}

\hfil

\noindent The two-point function in $n$ dimensions is:
\begin{equation}
    I_{2}^{n} \left( s \right) \ = \ \dsp \frac{1}{\epsilon (1-2\epsilon)} \left( -s \right)
    ^{-\epsilon} \ = \ \frac{1}{\epsilon} - \ln \left( -s \right) +2 + O\left( \epsilon
    \right),
\end{equation}
\noindent and in $n+2$ dimensions :
\begin{equation}
    I_{2}^{n+2} \left( s \right) \ = \ \dsp -\frac{1}{2\epsilon (1-2\epsilon)(3-2\epsilon)} \left( -s \right)
    ^{1-\epsilon}.
\end{equation}

\hfil

\subsection{One-external-massive-leg three-point Function.}

\begin{equation*}
    \parbox{2.5cm}{\includegraphics[width=2.5cm]{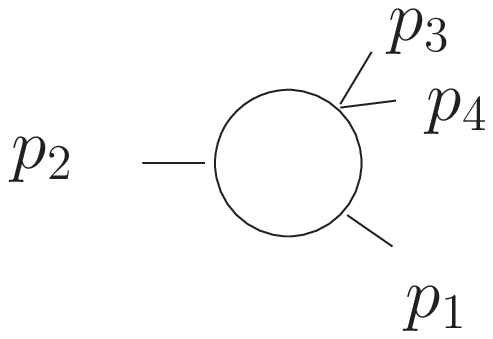}}
    \quad \quad \quad \quad \quad \quad \quad \quad
    s = s_{34}
\end{equation*}

\begin{align}
{{\cal S}_{3,1}} =
\begin{pmatrix}
    0 & 0 & s \\
    0 & 0 & 0   \\
    s & 0 & 0
\end{pmatrix}.
\end{align}

\noindent The determinants are given by:
\begin{align}
    \det \left( {\cal S}_{3,1} \right) & = 0  &\\
    \det \left( G_{3,1} \right) & = - s^{2}. &
\end{align}

\hfil

\noindent This one-external-massive-leg three-point function in
$n$ dimensions is:
\begin{equation}
    I_{3}^{n} \left( s \right) \ = \ \dsp \frac{1}{\epsilon ^{2}} \frac{\left( -s
    \right)^{-\epsilon}}{s} \ = \ \frac{1}{s} \left( \frac{1}{\epsilon^{2}} - \ln \left( -s \right)
    + \frac{\ln \left( -s \right) ^{2}}{2} \right)+ O\left( \epsilon \right),
\end{equation}
\noindent and in $n+2$ dimensions :
\begin{equation}
    I_{3}^{n+2} \left( s \right) \ = \ \dsp \frac{1}{2 \epsilon (1-\epsilon)(1-2\epsilon)} \left( -s
    \right)^{-\epsilon}.
\end{equation}

\hfil

\subsection{The two-external-massive-leg three-point Function}

\begin{equation*}
    \parbox{2.5cm}{\includegraphics[width=2.5cm]{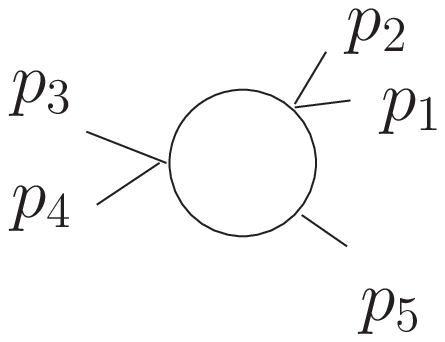}}
    \quad \quad \quad \quad \quad \quad \quad \quad
    \left\{\begin{array}{l}
        m_{1}^{2} = s_{12} \\
        m_{2}^{2} = s_{34}
    \end{array} \right.
\end{equation*}
\begin{align}
{{\cal S}_{3,2}} =
\begin{pmatrix}
    0 & m_{2}^{2} & m_{1}^{2} \\
    m_{2}^{2} & 0 & 0   \\
    m_{1}^{2} & 0 & 0
\end{pmatrix}.
\end{align}

\noindent The determinants are given by:
\begin{align}
    \det \left( {\cal S}_{3,2} \right) & \ = \ 0  &\\
    \det \left( G_{3,2} \right) & \ = \ -\left(m_{1}^{2} - m_{2}^{2}\right)^{2}. &
\end{align}

\noindent The two-external-massive-leg three-point function in $n$
dimensions is:
\begin{equation}
    I_{3}^{n} \left( m_{1}^{2},m_{2}^{2} \right) \ = \ \dsp \frac{1}{\epsilon ^{2}} \frac{\left(
    -m_{1}^{2}\right)^{-\epsilon}-\left(
    -m_{2}^{2}\right)^{-\epsilon}}{m_{1}^{2}-m_{2}^{2}}.
\end{equation}

\hfil

\subsection{The three-external-massive-leg three-point Function}

\begin{align*}
    \parbox{2.5cm}{\includegraphics[width=2.5cm]{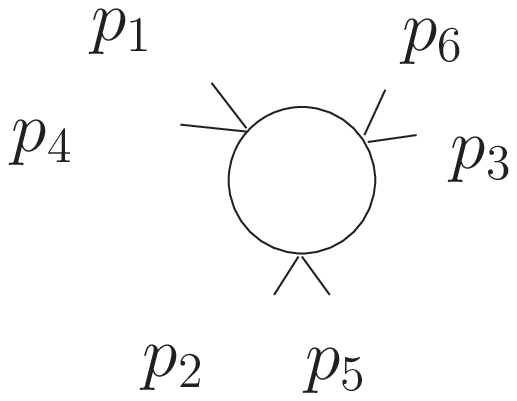}}
    \quad \quad \quad \quad \quad \quad \quad \quad
    \left\{\begin{array}{l}
        m_{1}^{2} = s_{14} \\
        m_{2}^{2} = s_{25} \\
        m_{3}^{2} = s_{36}
    \end{array} \right.
\end{align*}

\begin{align}
{{\cal S}_{3,3}} =
\begin{pmatrix}
    0 & m_{2}^{2} & m_{1}^{2} \\
    m_{2}^{2} & 0 & m_{3}^{2}   \\
    m_{1}^{2} & m_{3}^{2} & 0
\end{pmatrix}.
\end{align}
\noindent The determinants are given by:
\begin{align}
    \det (G_{3,3}) & = m_{1}^{2}m_{2}^{2}-\left( m_{1}.m_{2}\right)^{2} = - \frac{\Delta}{4}&\\
    \det ({\cal S}_{3,3}) & = 2m_{1}^{2}m_{2}^{2}m_{3}^{2}. &
\end{align}
\noindent The three-external-massive-leg three-point function in
$n$ dimensions is \cite{Binoth:2001vm}:
\begin{align}
    I_{3}^{n} \left( m_{1}^{2},m_{2}^{2},m_{3}^{2} \right) = &
    \dsp \frac{1}{\sqrt{\Delta}} \left\{ \left ( 2 \textrm{Li}_{2} \left( 1-
    \frac{1}{y_{2}}\right) + 2 \textrm{Li}_{2} \left( 1-
    \frac{1}{x_{2}}\right) + \frac{\pi^{2}}{3} \right) \right.& \nonumber \\
    & \left.\dsp +\frac{1}{2}\left ( \ln^{2}\left( \frac{x_{1}}{y_{1}}
    \right)+ \ln^{2}\left( \frac{x_{2}}{y_{2}} \right) + \ln^{2}\left( \frac{x_{2}}{y_{1}}
    \right) - \ln^{2}\left( \frac{x_{1}}{y_{2}} \right) \right) \right\}, &
    \label{3p3m}
\end{align}

\noindent where:

\begin{align}
    x_{1,2} & = \dsp \frac{m_{1}^{2} + m_{2}^{2} - m_{3}^{2} \pm \sqrt{\Delta} }{ 2
    m_{1}^{2}}& \\
    y_{1,2} & = \dsp \frac{m_{1}^{2} - m_{2}^{2} + m_{3}^{2} \pm \sqrt{\Delta} }{ 2
    m_{1}^{2}}& \\
    \Delta & = \dsp m_{1}^4 + m_{2}^4 + m_{3}^4 -2 m_{1}^{2} m_{2}^{2} -2 m_{1}^{2}m_{3}^{2} -2
    m_{2}^{2}m_{3}^{2} -i \ \textrm{sign}(m_{1}^{2}) \ \epsilon . &
\end{align}

\noindent The formula $\left(\ref{3p3m} \right)$ is available
every where thanks to the small imaginary part $i \ \epsilon $~:
\begin{eqnarray}
    \sqrt{\Delta \pm i \epsilon}  & = & \left\{ \begin{array}{l}
    \sqrt{\Delta} \pm i \epsilon \, , \;
    \Delta \geq 0 \\
    \pm i \, \sqrt{-\Delta} \, , \; \Delta \leq 0
    \end{array} \right.
\end{eqnarray}

\hfil

\subsection{The zero-external-massive-leg four-point Function}

\begin{align*}
    \parbox{2.5cm}{\includegraphics[width=2.5cm]{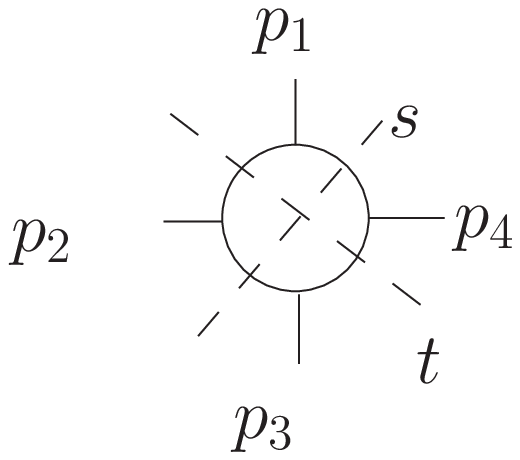}}
    \quad \quad \quad \quad \quad \quad \quad \quad
    \left\{\begin{array}{l}
    s = s_{12} \\
    t = s_{14} \\
    u = s_{13}
    \end{array} \right.
\end{align*}

\begin{align}
{{\cal S}_{4,0}} =
\begin{pmatrix}
    0 & 0 & s & 0 \\
    0 & 0 & 0 & t \\
    s & 0 & 0 & 0 \\
    0 & t & 0 & 0
\end{pmatrix}.
\end{align}

\noindent The determinants are given by:
\begin{align}
    \det  \left( G_{4,0} \right) & \ = \ -2st(s+t) \ = \ 2stu & \\
    \det  \left( {\cal S}_{4,0} \right) & \ = \ t^{2}s^{2} \ = \ \langle 24342 \rangle ^{2}. &
\end{align}

\hfil

\noindent The zero-external-massive-leg four-point function in $n$
dimensions is \cite{Binoth:2001vm} :
\begin{equation}
    I_{4,0}^{n} \left( s,t \right) \ = \ \dsp \frac{2}{st} \frac{1}{\epsilon
    ^{2}}\left\{ (-s)^{-\epsilon} + (-t)^{-\epsilon} \right\} -\frac{2}{ st} F_{0}
    (s,t),
\end{equation}
\noindent where:
\begin{equation}
    F_{0} (s,t) \ = \ \dsp \frac{1}{2}\left\{  \ln ^{2} \left( \frac{s}{t} \right) + \pi^{2}
    \right\}.
\end{equation}
\noindent And in $n+2$ dimensions, it is:
\begin{equation}
    I_{4,0}^{n+2} \left( s,t \right) \ =  \ \frac{F_{0} (s,t)}{u(n-3)}.
\end{equation}

\hfil

\subsection{The one-external-massive-leg four-point Function}

\begin{equation*}
\parbox{2.5cm}{\includegraphics[width=2.5cm]{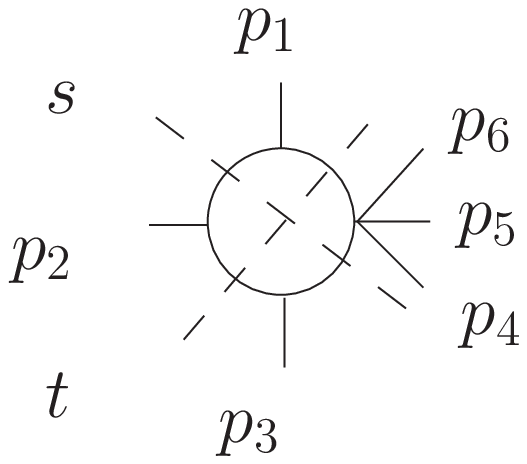}}
\quad \quad \quad \quad \quad \quad \quad \quad
\left\{\begin{array}{l}
    s = s_{12} \\
    t = s_{23} \\
    u = s_{13} \\
    m^{2} = s_{456}
\end{array} \right.
\end{equation*}

\begin{equation}
{{\cal S}_{4,1}} =
\begin{pmatrix}
    0 & m_{1}^{2} & s & 0 \\
    m_{1}^{2} & 0 & 0 & t  \\
    s & 0 & 0 & 0   \\
    0 & t & 0 & 0
\end{pmatrix}.
\end{equation}

\noindent The determinants are given by:
\begin{align}
    \det (G_{4,1}) & \ = \ -2st\left(s+t-m^{2}\right) \ = \ 2stu \\
    \det  (S_{4,1}) & \ = \ (st)^{2} =  \langle 1m3m1 \rangle ^{2}.
\end{align}

\noindent The one-external-massive-leg four-point function in $n$
dimensions is \cite{Binoth:2001vm}:

\begin{align}
    I_{4,1}^{n} \left( s,t,m^{2} \right) = & \dsp \ \  \frac{1}{st \epsilon
    ^{2}}\left\{ \left( (-s)^{-\epsilon} + (-t)^{-\epsilon} \right)
    + \left( (-s)^{-\epsilon} - (-m^{2})^{-\epsilon} \right)
    + \left( (-t)^{-\epsilon} - (-m^{2})^{-\epsilon} \right)
    \right\} & \nonumber \\
    & \dsp - \frac{2}{st} F_{1} \left(s,t,m^{2} \right), &
\end{align}
\noindent where:
\begin{align}
    F_{1} \left(s,t,m^{2} \right)& \dsp = \textrm{Li}_{2} \left(1 - \frac{m^{2}}{s}\right) + \textrm{Li}_{2} \left(1 -
    \frac{m^{2}}{t} \right) - \textrm{Li}_{2} \left(- \frac{s}{t} \right) - \textrm{Li}_{2} \left(- \frac{t}{s} \right) & \\
    & = F_{0} (s,t) + \left\{ \textrm{Li}_{2} \left(1 - \frac{m^{2}}{s} \right) + \textrm{Li}_{2} \left(1 - \frac{m^{2}}{t} \right) -
    \frac{\pi^{2}}{3} \right\}. &
\end{align}
\noindent And in $n+2$ dimensions, it is:
\begin{equation}
    I_{4,1}^{n+2} \left( s,t,m^{2} \right) \ = \ \frac{F_{1} (s,t,m^{2})}{u(n-3)}.
\end{equation}

\hfil

\subsection{The two-adjacent-external-massive-leg four-point Function}

\begin{equation*}
\parbox{2.5cm}{\includegraphics[width=2.5cm]{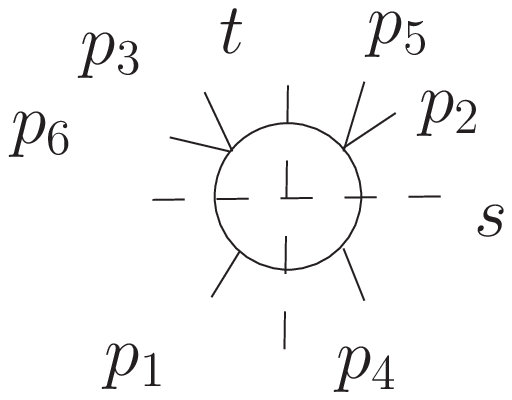}}
\quad \quad \quad \quad \quad \quad \quad \quad
\left\{\begin{array}{l}
    s = s_{14} \\
    t = s_{425} \\
    u = s_{125} \\
    m_{1}^{2} = s_{25} \\
    m_{2}^{2} = s_{36}
\end{array} \right.
\end{equation*}

\begin{align}
{{\cal S}_{4,2A}} =
\begin{pmatrix}
    0 & m_{1}^{2} & s & 0 \\
    m_{1}^{2} & 0 & m_{2}^{2} & t  \\
    s & m_{2}^{2} & 0 & 0   \\
    0 & t & 0 & 0
\end{pmatrix}.
\end{align}

\noindent The determinants are given by:
\begin{align}
     \det (G_{4,2A}) & = \ -2s\left(m_{1}^{2}m_{2}^{2} -t(m_{1}^{2}+m_{2}^{2}-s-t)\right) = \ -2s\langle 1m_{1}4m_{2}1
     \rangle&\\
     \det  (S_{4,2A}) & = \ (st)^{2} &
\end{align}

\noindent The two-adjacent-external-massive-leg four-point
Function in $n$ dimensions is:
\begin{align}
    I_{4,2A}^{n} \left( s,t,m_{1}^{2},m_{2}^{2} \right) = & \dsp \ \ \frac{1}{(st) \epsilon
    ^{2}}\left\{ (-s)^{-\epsilon}
    + \left( (-t)^{-\epsilon} - (-m_{1}^{2})^{-\epsilon} \right)
    + \left( (-t)^{-\epsilon} - (-m_{2}^{2})^{-\epsilon} \right) \right\} & \nonumber \\
    & \dsp -\frac{2}{st} F_{2A} \left(s,t,m_{1}^{2},m_{2}^{2} \right), &
\end{align}
\noindent where:
\begin{align}
    F_{2A} \left(s,t,m_{1}^{2},m_{2}^{2} \right) & = \dsp \textrm{Li}_{2} \left(1 - \frac{m_{1}^{2}}{t} \right) + \textrm{Li}_{2} \left(1 - \frac{m_{2}^{2}}{t}\right)
    + \frac{1}{2} \ln \left( \frac{s}{t} \right) \ln \left( \frac{m_{2}^{2}}{t} \right)+
    \frac{1}{2} \ln \left( \frac{s}{m_{2}^{2}} \right) \ln \left( \frac{m_{1}^{2}}{t} \right). &
\end{align}
\noindent And in $n+2$, it is:
\begin{align}
    I_{4,2A}^{n+2} \left( s,t,m_{1}^{2},m_{2}^{2} \right) = & \ \ \frac{t}{\left(tu -m_{1}^{2}m_{2}^{2} \right)(n-3)}F_{2A}
    (s,t,m_{1}^{2},m_{2}^{2}) &\nonumber \\
    & - \frac{2m_{1}^{2}m_{1}^{2} +t\left( s-m_{1}^{2}-m_{2}^{2}\right)}{2(n-3)\left(tu -m_{1}^{2}m_{2}^{2}
    \right)}I_{3}^{n}\left(m_{1}^{2},m_{2}^{2},m_{3}^{2} \right). &
\end{align}

\hfil

\subsection{The two-opposite-external-massive-leg four-point Function}

\begin{equation*}
\parbox{2.5cm}{\includegraphics[width=2.5cm]{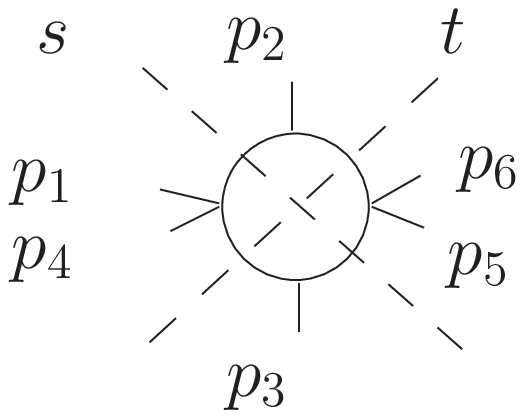}}
\quad \quad \quad \quad \quad \quad \quad \quad
\left\{\begin{array}{l}
    s = s_{143} \\
    t = s_{243} \\
    u = s_{23} \\
    m_{1}^{2} = s_{14} \\
    m_{2}^{2} = s_{56}
\end{array} \right.
\end{equation*}

\begin{align}
{{\cal S}_{4,2B}} =
\begin{pmatrix}
    0 & m_{1}^{2} & s & 0 \\
    m_{1}^{2} & 0 & 0 & t  \\
    s & 0 & 0 & m_{2}^{2}   \\
    0 & t & m_{2}^{2} & 0
\end{pmatrix}.
\end{align}
\noindent The determinants are given by:
\begin{align}
    \det (G_{4,2B}) & = -2\left(m_{1}^{2}m_{2}^{2}-st \right) \left(m_{1}^{2}+m_{2}^{2}-s-t\right) = 2u\left(st-m_{1}^{2}m_{2}^{2}\right) &\\
    \det (S_{4,2B}) & = \left( st -m_{1}^{2}m_{2}^{2} \right)^{2} = \langle 2m_{1}3m_{1}2 \rangle ^{2} = \langle 2m_{2}3m_{2}2 \rangle ^{2}. &
\end{align}
\noindent The two-opposite-external-massive-leg four-point
function in $n$ dimensions is \cite{Binoth:2001vm}:
\begin{align}
    I_{4,2B}^{n} \left( s,t,m_{1}^{2},m_{2}^{2} \right) = & \dsp \frac{1}{(st-m_{1}^{2}m_{2}^{2}) \epsilon
    ^{2}}\left\{ \left( (-s)^{-\epsilon} - (-m_{1}^{2})^{-\epsilon} \right)
    + \left( (-s)^{-\epsilon} - (-m_{2}^{2})^{-\epsilon} \right)
    \right\} & \nonumber \\
    & + \dsp \frac{1}{(st-m_{1}^{2}m_{2}^{2}) \epsilon
    ^{2}}\left\{ \left( (-t)^{-\epsilon} - (-m_{1}^{2})^{-\epsilon} \right) + \left( (-t)^{-\epsilon} - (-m_{2}^{2})^{-\epsilon} \right)
    \right\} & \nonumber \\
    &  \dsp -\frac{2}{st - m_{1}^{2}m_{2}^{2}} F_{2B} \left(s,t,m_{1}^{2},m_{2}^{2} \right), &
\end{align}
\noindent where:
\begin{align}
    F_{2B} \left(s,t,m_{1}^{2},m_{2}^{2} \right) = & \dsp  \  - \textrm{Li}_{2} \left(1 - \frac{m_{1}^{2}m_{2}^{2}}{st}
    \right) + \textrm{Li}_{2} \left(1 - \frac{m_{1}^{2}}{s} \right) & \nonumber \\
    & \dsp + \textrm{Li}_{2} \left(1 - \frac{m_{2}^{2}}{s} \right)
    + \textrm{Li}_{2} \left(1 - \frac{m_{1}^{2}}{t} \right) + \textrm{Li}_{2} \left(1 - \frac{m_{2}^{2}}{t} \right) +  \frac{1}{2} \ln ^{2} \left( \frac{s}{t}\right) & \\
    = & \dsp \ F_{1} \left(s,t,m_{1}^{2} \right) + F_{1} \left(s,t,m_{2}^{2}
    \right)- F_{0} (s,t) - \left\{ \textrm{Li}_{2} \left( 1 - \frac{m_{1}^{2}m_{2}^{2}}{st} \right)- \frac{\pi^{2}}{6} \right\}. &
\end{align}
\noindent And in $n+2$ dimensions, it is:
\begin{equation}
    I_{4,2B}^{n+2} \left( s,t,m_{1}^{2},m_{2}^{2} \right) \ = \ \frac{F_{2B}
    (s,t,m_{1}^{2},m_{2}^{2})}{u(n-3)}.
\end{equation}

\hfil

\subsection{The three-external-massive-leg four-point Function}

\begin{equation*}
\parbox{2.5cm}{\includegraphics[width=2.5cm]{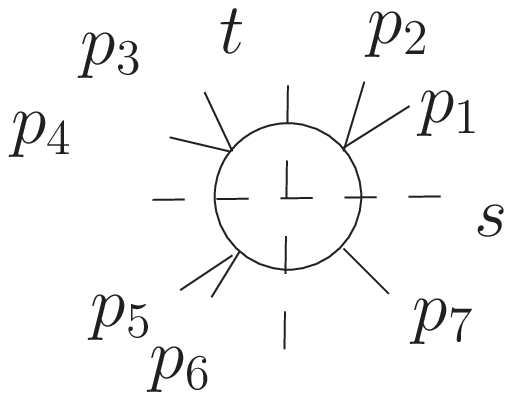}}
\quad \quad \quad \quad \quad \quad \quad \quad
\left\{\begin{array}{l}
    s = s_{1234} \\
    t = s_{712} \\
    u = s_{1256} \\
    m_{1}^{2} = s_{12} \\
    m_{2}^{2} = s_{34} \\
    m_{3}^{2} = s_{56}
\end{array} \right.
\end{equation*}
\begin{align}
{{\cal S}_{4,3}} =
\begin{pmatrix}
    0 & m_{1}^{2} & s & 0 \\
    m_{1}^{2} & 0 & m_{2}^{2} & t  \\
    s & m_{2}^{2} & 0 & m_{3}^{2}   \\
    0 & t & m_{3}^{2} & 0
\end{pmatrix}.
\end{align}
\noindent The determinant is:
\begin{equation}
    \det ({\cal S}_{4,3}) \ = \ \left( st - m_{1}^{2}m_{3}^{2} \right)^{2} = \langle 7m_{1}m_{2}m_{1}7 \rangle ^{2} = \langle 7m_{3}m_{2}m_{3}7 \rangle
    ^{2}.
\end{equation}

\noindent The three-external-massive-leg four-point function in
$n$ dimensions is:
\begin{align}
    I_{4}^{n} \left( s,t,m_{1}^{2},m_{2}^{2},m_{3}^{2} \right) = & \dsp \frac{1}{(st-m_{1}^{2}m_{3}^{2}) \epsilon
    ^{2}}\left\{ (-s)^{-\epsilon} + (-t)^{-\epsilon} - (-m_{1}^{2})^{-\epsilon}
    - (-m_{2}^{2})^{-\epsilon}- (-m_{3}^{2})^{-\epsilon} \right\} & \nonumber \\
    &  \dsp -\frac{2}{st - m_{1}^{2}m_{3}^{2}} F_{3} \left(s,t,m_{1}^{2},m_{2}^{2},m_{3}^{2} \right), &
\end{align}
\noindent where:
\begin{align}
    F_{3} \left(s,t,m_{1}^{2},m_{2}^{2},m_{3}^{2} \right) = & \dsp
    - \frac{1}{2} \ln \left( \frac{s}{m_{1}^{2}} \right) \ln \left(
    \frac{s}{m_{2}^{2}} \right) - \frac{1}{2} \ln \left( \frac{t}{m_{3}^{2}} \right) \ln \left(
    \frac{t}{m_{2}^{2}} \right) + \frac{1}{2} \ln ^{2} \left(
    \frac{s}{t} \right) & \nonumber \\
    & + \textrm{Li}_{2} \left(1 - \frac{m_{1}^{2}}{s} \right) + \textrm{Li}_{2} \left(1 - \frac{m_{3}^{2}}{t}
    \right)- \textrm{Li}_{2} \left(1 - \frac{m_{1}^{2}m_{3}^{2}}{st}
    \right).&
\end{align}

\subsection{The four-external-massive-leg four-point Function}

\begin{equation*}
\parbox{2.5cm}{\includegraphics[width=2.5cm]{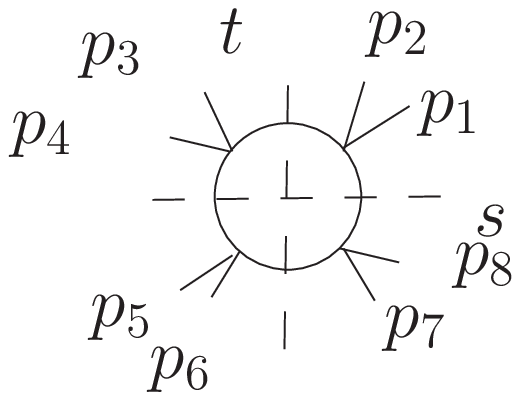}}
\quad \quad \quad \quad \quad \quad \quad \quad
\left\{\begin{array}{l}
    s = s_{1234} \\
    t = s_{7812} \\
    u = s_{1256} \\
    m_{1}^{2} = s_{12} \\
    m_{2}^{2} = s_{34} \\
    m_{3}^{2} = s_{56} \\
    m_{4}^{2} = s_{78}
\end{array} \right.
\end{equation*}
\begin{align}
{{\cal S}_{4,4}} =
\begin{pmatrix}
    0 & m_{1}^{2} & s & m_{4}^{2} \\
    m_{1}^{2} & 0 & m_{2}^{2} & t  \\
    s & m_{2}^{2} & 0 & m_{3}^{2}   \\
    m_{4}^{2} & t & m_{3}^{2} & 0
\end{pmatrix}.
\end{align}

\noindent The determinant is given by:
\begin{equation}
    \det ({\cal S}_{4,4}) \ = \ \left( st -m_{1}^{2}m_{3}^{2} -m_{2}^{2}m_{4}^{4}\right)^{2}
    -4m_{1}^{2}m_{3}^{2}m_{2}^{2}m_{4}^{4}.
\end{equation}
\noindent The four-external-massive-leg four-point Function in $n$
dimensions is:
\begin{align}
    I_{4}^{n} \left( s,t,m_{1}^{2},m_{2}^{2},m_{3}^{2},m_{4}^{2} \right) \ = \ \frac{1}
    {\left(m_{1}^{2}+m_{2}^{2} \right)^{2}\left(m_{2}^{2}+m_{3}^{2}\right)^{2} \rho}F_{4} \left(s,t,m_{1}^{2},m_{2}^{2},m_{3}^{2},m_{4}^{2} \right),&
\end{align}
\noindent where:
\begin{align}
    F_{4} \left(s,t,m_{1}^{2},m_{2}^{2},m_{3}^{2},m_{4}^{2} \right) \ =  \
    \frac{1}{2} & \dsp \left\{  -\textrm{Li}_{2} \left( \frac{1-\lambda_{1}+\lambda_{2}+\rho}{2}
    \right)+ -\textrm{Li}_{2} \left( \frac{1-\lambda_{1}+\lambda_{2}-\rho}{2}
    \right)\right. & \nonumber \\
    & \ \ -\textrm{Li}_{2} \left( \frac{1-\lambda_{1}-\lambda_{2}-\rho}{2\lambda_{1}}
    \right) + \textrm{Li}_{2} \left( \frac{1-\lambda_{1}-\lambda_{2}+\rho}{2\lambda_{1}}
    \right) & \nonumber \\
    &\left. \ \  - \frac{1}{2} \ln \left( \frac{\lambda_{1}}{\lambda_{2}^{2}} \right)
    \ln \left( \frac{1+\lambda_{1}-\lambda_{2}+\rho}{1+\lambda_{1}-\lambda_{2}-\rho}\right)\right\}, &
\end{align}
\noindent and:
\begin{align}
    \rho & \ = \ \sqrt{1-2\lambda_{1}-2\lambda_{2}+\lambda_{1}^{2}-2\lambda_{1}\lambda_{2}+\lambda_{2}^{2}}&
    \\
    \lambda_{1} & \ = \ \frac{m_{1}^{2}m_{3}^{2}}{\left(m_{1}^{2}+m_{2}^{2} \right)^{2}\left(m_{2}^{2}+m_{3}^{2}\right)^{2}} &\\
    \lambda_{2} & \ = \ \frac{m_{2}^{2}m_{4}^{2}}{\left(m_{1}^{2}+m_{2}^{2} \right)^{2}\left(m_{2}^{2}+m_{3}^{2}\right)^{2}}. &
\end{align}

\newpage

\section{Recall of the formalism introduced by Forde \cite{Forde:2007mi}.} \label{Forderecall}

\hfil

\noindent We have:
\begin{equation}
    \left\{ \begin{array}{l}
    \dsp K_{3}^{b} \ = \ \frac{p_{3} - \left( p_{4}^{2} / \gamma \right) p_{4}}{1- \left( p_{3}^{2}p_{4}^{2}/ \gamma^{2}\right)} \\
    \dsp K_{4}^{b} \ = \ \frac{p_{4} - \left( p_{4}^{2} / \gamma \right) p_{3}}{1- \left( p_{3}^{2}p_{4}^{2}/
    \gamma^{2}\right)},
    \end{array} \right.
\end{equation}

\noindent and

\begin{equation}
    \left\{ \begin{array}{l}
    \dsp \alpha_{03} \ = \ \frac{p_{3}^{2}\left( \gamma - p_{4}^{2}\right)}{\gamma^{2}- p_{3}^{2}p_{4}^{2}} \\
    \dsp \alpha_{04} \ = \ \frac{p_{4}^{2}\left( \gamma - p_{3}^{2}\right)}{\gamma^{2}- p_{3}^{2}p_{4}^{2}},
    \end{array} \right.
\end{equation}

\noindent with $ \dsp \gamma \ = \ p_{3}.p_{4} \pm \sqrt{\left(
p_{3}.p_{4}\right)^{2}-p_{3}^{2}p_{4}^{2}}$.

\end{appendix}

\end{document}